\input phyzzx
\hsize=417pt 
\sequentialequations
\Pubnum={ EDO-EP-10}
\date={ \hfill March 1997}
\titlepage
\vskip 32pt
\title{ Quantum Instability of Black Hole Singularity in Three 
Dimensions }
\author{Ichiro Oda \footnote\dag {E-mail address: 
sjk13904@mgw.shijokyo.or.jp}}
\vskip 12pt
\address{ Edogawa University,                                
          474 Komaki, Nagareyama City,                        
          Chiba 270-01, JAPAN     }                          
%
%
%
%
%
\abstract{ We study an internal structure of (2+1)-dimensional 
black hole with the neutral scalar matter in the spherically 
symmetric geometry by using a quantum theory of gravity which holds in 
the both vicinities of the singularity and the apparent horizon. A 
special attention is paid to the quantum-mechanical behavior of the 
singularity in the black hole. We solve analytically the Wheeler-DeWitt 
equation of a minisuperspace model where the ingoing Vaidya metric is 
used as a simple model representing a dynamical black hole. The wave 
function obtained in this way leads to interesting physical phenomena 
such as the quantum instability of singularity and the Hawking radiation.
It is also pointed out a similarity between the singularity in 
(2+1)-dimensional black hole and the inner Cauchy horizon in 
(3+1)-dimensional Reissner-Nordstrom charged black hole. } 
\endpage
%
%
%

\def\sp(#1){\noalign{\vskip #1pt}}

%
%
%
%
%
\topskip 30pt
\par
There is no doubt that understanding the unsolved problems associated 
with quantum black holes gives us some insight into a construction of a 
theory of quantum gravity. These open problems include the information 
loss paradox, the fate of the endpoint of black hole evaporation, and the 
statistical mechanical origin of black hole entropy e.t.c. [1]

It is an interesting idea that the physical features of a black 
hole can be attributed to the properties of the singularity and/or the 
event horizon. 
Actually, in those days, Thorne et al. [2] have to a certain extent 
developed this idea as the Membrane Paradigm where the stretched horizon, 
in the replacement of the event horizon, plays a fundamental role in 
describing various properties of a black hole, afterward being 
rediscovered by a different viewpoint [3, 4]. One of motivations in this 
article is not only to push forward with this idea but also to extend it 
to a direction that the quantum-mechanical properties of the singularity 
as well as   
the horizon essentially determine an overall physical behavior of quantum 
black holes.

However, it seems to be difficult to construct a quantum theory of
black holes. One of the standard approaches adopted so far is to reduce 
an infinitely many dynamical degrees of freedom to be finite ones by 
considering the simpler models, the minisuperspace models [5], and then 
construct a quantum mechanics of this system with only finitely many 
physical degrees of freedom. Though some minisuperspace models are surely 
effective and soluble, they are not so useful owing to a wildly singular 
behavior at the curvature singularity in applying them for study of a 
quantum mechanics near the singularity of black holes in four dimensions 
.

Recently, motivated by an interesting idea of Tomimatsu [6], in a series 
of papers we have developed a minisuperspace model of quantum black holes 
[7-9]. Our attentions were paid to the quantum-mechanical properties of  
black holes only in the vicinity of the apparent horizons. As mentioned 
above, our philosophy is that the black hole physics can be in 
essence understood 
in terms of a quantum theory holding near both the singularity 
and the horizon, thus our purpose has been reached only halfway so far.

It was recently discovered that there are black hole solutions in the 
anti-de Sitter spacetime in three dimensions, and clarified that the 
singularity is not the curvature singularity but has a rather mild 
character [10]. Thus it might be possible to argue various quantum 
aspects of the singularity  by means of the (2+1)-dimensional black hole. 
This   
is indeed the case as seen later. In particular, we will see that the 
singularity is quantum-mechanically unstable and becomes the curvature 
singularity under a slight matter perturbation.

We start with the three dimensional action that is of the form
$$ \eqalign{ \sp(2.0)
S = \int \ d^3 x \sqrt{-^{(3)}g} \ \bigl[ {1 \over {16 \pi G}} \bigl( 
{}^{(3)}R + {2 \over l^2} \bigr) - {1 \over 8 \pi} {}^{(3)}g^{\mu\nu} 
\partial_{\mu} \Phi \partial_{\nu} \Phi \bigr],
\cr
\sp(3.0)} \eqno(1)$$
where the cosmological constant $\Lambda$ is related to the scale 
parameter $l$ by $\Lambda = - {1 \over l^2}$, and 
$\Phi$ is a real scalar field.  To exhibit explicitly the three 
dimensional character we put the suffix $(3)$ 
in front of the metric tensor and the curvature scalar. 
In the previous work [9], we have considered the more general matter 
contents, but we will confine ourselves to the action (1) for simplicity. 
We will follow the conventions adopted in the MTW textbook 
[11] and use the natural units $G = \hbar = c = 1$. The Greek indices 
$\mu, \nu, ...$ take 0, 1 and 2, and the Latin indices 
$a, b, ...$ take 0 and 1. 

Let us make the most general spherically symmetric reduction for 
the metric 
$$ \eqalign{ \sp(2.0)
ds^2 &= {}^{(3)}g_{\mu\nu} dx^{\mu} dx^{\nu},
\cr
     &= g_{ab} dx^a dx^b + \phi^2 d\theta^2, 
\cr
\sp(3.0)} \eqno(2)$$
with the two dimensional metric $g_{ab}$ and the radial function $\phi$ 
being the function of only the two dimensional coordinates $x^a$, 
and the angular variable $\theta$ ranging from 0 to $2 \pi$.
An integration over the angular variable $\theta$ after substituting (2) 
into (1) yields the following effective action in two dimensions:
$$ \eqalign { \sp(2.0)
S &= {1 \over 8} \int \ d^2 x \sqrt{-g} \  \phi \ \bigl( R + {2 \over l^2} 
\bigr) - {1 \over 4} \int \ d^2 x \sqrt{-g} \ \phi \  g^{ab} \partial_a 
\Phi \partial_b \Phi. 
\cr
\sp(3.0)} \eqno(3)$$
It was pointed out that the gravitational sector in this action has
a curious feature that it is cast into the topological BF theory, which 
is a peculiar feature in the s-wave reduction from three to two 
dimensions [9].

For later convenience, let us derive the equations of motion arising from 
the action (1):
$$ \eqalign{ \sp(2.0)
R + {2 \over l^2}  - 2  g^{ab} \partial_a \Phi \partial_b \Phi = 0, 
\cr
\sp(3.0)} \eqno(4)$$
$$ \eqalign{ \sp(2.0)
\nabla_a \nabla_b \phi - g_{ab} \nabla_c \nabla^c \phi +  g_{ab} {1 
\over l^2} \phi + 2 \phi \ (\partial_a \Phi \partial_b \Phi - {1 \over 2} 
g_{ab} \partial_c \Phi \partial^c \Phi) = 0,
\cr
\sp(3.0)} \eqno(5)$$
$$ \eqalign{ \sp(2.0)
\partial_a (\sqrt{-g} \ \phi \ g^{ab} \ \partial_b \Phi) = 0.
\cr
\sp(3.0)} \eqno(6)$$

Since we would like to discuss the canonical formalism of the action 
(1), particularly, in the vicinity of the black hole singularity, our 
interest lies in the interior of the (2+1)-dimensional black hole, 
consisting of region bounded between the singularity and the horizon. The 
important point here is that in the interior the role of time $x^0$ and 
space $x^1$ is exchanged owing to the signature structure of the metric 
tensor so that one has to foliate the interior of a black hole by a 
family of spacelike hypersurfaces, for instance, $x^1 = r = const$. This 
canonical formalism has been discussed in detail in ref.[7] in the case 
of the four dimensional gravity, so we will write only the results 
adapted for the present purpose.

By using a proper ADM splitting of (1+1)-dimensional spacetime
$$ \eqalign{ \sp(2.0)
g_{ab} = \left(\matrix{ \gamma  & \alpha \cr
              \alpha & {\alpha^2 \over \gamma} - \beta^2 \cr} \right),
\cr
\sp(3.0)} \eqno(7)$$
the normal unit vector orthogonal to the hypersurface 
$x^1 = const$ 
$$ \eqalign{ \sp(2.0)
n^a = ({\alpha \over {\beta \gamma}}, \ - {1 \over \beta}),
\cr
\sp(3.0)} \eqno(8)$$
and the trace of the extrinsic curvature
$$ \eqalign{ \sp(2.0)
K &= {1 \over \sqrt{-g}} \partial_a (\sqrt{-g} \ n^a) , 
\cr
&= - {\gamma^\prime \over {2\beta\gamma}} + {\dot \alpha \over 
{\beta\gamma}} - {\alpha \over {2\beta\gamma^2}} \dot \gamma,
\cr
\sp(3.0)} \eqno(9)$$
the action (3) can be written 
$$ \eqalign{ \sp(2.0)
S &=\int d^2x \ {1 \over 4} \beta \sqrt{\gamma} \ 
\bigl( - K n^a \partial_a \phi + {\dot \beta \over {\beta\gamma}}
\dot \phi + {1 \over l^2} \phi \bigr) 
\cr
&\qquad+ \int d^2x \ {1 \over 4} \beta \sqrt{\gamma} \ \phi \ 
\bigl[ ( n^a \partial_a \Phi )^2 
- {1 \over \gamma} (\partial_0 \Phi)^2 \bigr], 
\cr
\sp(3.0)} \eqno(10)$$
where ${\partial \over {\partial x^0}} = \partial_0$ and ${\partial \over 
{\partial x^1}} = \partial_1$ are also denoted by an overdot and a prime, 
respectively. The canonical conjugate momenta can be read off from the 
action (10). They are
$$ \eqalign{ \sp(2.0)
p_{\Phi} = - {\sqrt{\gamma} \over 2} \phi \ n^a \partial_a \Phi,
\cr
\sp(3.0)} \eqno(11)$$
$$ \eqalign{ \sp(2.0)
p_{\phi} = {\sqrt{\gamma} \over 4}  K,
\cr
\sp(3.0)} \eqno(12)$$
$$ \eqalign{ \sp(2.0)
p_{\gamma} =  {1 \over 8 \sqrt \gamma} n^a \partial_a \phi.
\cr
\sp(3.0)} \eqno(13)$$
Then the Hamiltonian can be calculated to be
$$ \eqalign{ \sp(2.0)
H = \int dx^0 (  \beta H_0 + \alpha H_1 ), 
\cr
\sp(3.0)} \eqno(14)$$
where the constraints $H_0$ and $H_1$ are explicitly given by
$$ \eqalign{ \sp(2.0)
H_0 = {1 \over {\sqrt{\gamma} \phi}} p_{\Phi}^2  - 8 \sqrt{\gamma} 
p_\phi p_\gamma + {1 \over 4} \partial_0 ( {\dot \phi \over 
\sqrt{\gamma}}) - {\sqrt{\gamma} \over 4} {1 \over l^2} \phi  
+ {\phi \over {4 \sqrt{\gamma}}} ( \partial_0 \Phi )^2 , 
\cr
\sp(3.0)} \eqno(15)$$
$$ \eqalign{ \sp(2.0)
H_1 = {1 \over \gamma} p_\Phi \partial_0 \Phi 
 + {1 \over \gamma} p_\phi \dot \phi - 2 \dot p_\gamma - {1 \over 
\gamma} p_\gamma \dot \gamma. 
\cr
\sp(3.0)} \eqno(16)$$

We now turn our attention to a canonical quantization when applied to the 
(2+1)-dimensional dynamical black hole. To begin with, let us introduce 
the two dimensional coordinates $x^a$ by 
$$ \eqalign{ \sp(2.0)
x^a = (x^0, x^1) = (v - r, r),
\cr
\sp(3.0)} \eqno(17)$$
where the advanced time coordinate is defined by $v = t + r^*$ with the 
tortoise coordinate $dr^* = {dr \over {-g_{00}}}$. In this coordinate 
system, we fix the gauge freedoms corresponding to the two dimensional 
reparametrization invariances by the gauge conditions
$$ \eqalign{ \sp(2.0)
g_{ab} &= \left(\matrix{ \gamma  & \alpha \cr
              \alpha & {\alpha^2 \over \gamma} - \beta^2 \cr} \right),
\cr
 &= \left(\matrix{ - ( - M + {r^2 \over l^2})   &  1 + M - {r^2 \over l^2}
               \cr
              1 + M - {r^2 \over l^2} & 2 + M - {r^2 \over l^2} \cr} 
              \right),
\cr
\sp(3.0)} \eqno(18)$$
where the black hole mass $M$ is the function of the two dimensional 
coordinates $x^a$, on the other hand, the scale parameter $l$ is a 
constant. Then the two dimensional line element takes a form of the 
Vaidya metric corresponding to the three dimensional black hole without 
rotation 
$$ \eqalign{ \sp(2.0)
ds^2 &= g_{ab} dx^a dx^b,
\cr
     &= - ( - M + {r^2 \over l^2}) dv^2 + 2 dv dr,
\cr
\sp(3.0)} \eqno(19)$$
which we will use as a model representing a dynamical black hole.

At this point it seems to be appropriate to make some comments on a 
canonical quantization of a different sort of minisuperspace model of a 
black hole in order to explain why we have to make use of the present 
rather general formalism to examine the properties of a dynamical black 
hole. Recall that now we take account of only the interior of a 
spherically symmetric black hole, for which we may choose the three 
dimensional line element to be 
$$ \eqalign{ \sp(2.0)
ds^2 = - N(r)^2 dr^2 + a(r)^2 dt^2 + \phi(r)^2 d\theta^2.
\cr
\sp(3.0)} \eqno(20)$$
This minisuperspace model has a dependency on only the r variable in the 
components of the metric tensor as a natural generalization of the 
solution satisfying the vacuum Einstein equation in three dimensions. To 
this minisuperspace model we can easily perform a canonical quantization 
without needing to appeal to the above general canonical 
formalism. Here let us briefly mention how to quantize the system (20). 
Without matter field, under the metric ansatz (20) the action (10) takes 
a form
$$ \eqalign{ \sp(2.0)
S = L \int_0 ^{r_{+}} dr \ \bigl( -{1 \over N} a^{\prime} \phi^{\prime} 
+ {1 \over l^2} N  a  \phi \bigr), 
\cr
\sp(3.0)} \eqno(21)$$
where we have defined to be $L = {1 \over 4} \int_{-T} ^T dt$ with the 
infrared cutoff $T$, and $r_{+}$ denotes the radius of the horizon. The 
canonical conjugate momenta corresponding to the dynamical variables $a$ 
and $\phi$ become
$$ \eqalign{ \sp(2.0)
\pi_a &= - {L \over N} \phi^{\prime} , 
\cr
\pi_{\phi} &= - {L \over N} a^{\prime} ,
\cr
\sp(3.0)} \eqno(22)$$
and the Hamiltonian becomes
$$ \eqalign{ \sp(2.0)
H = - {N \over L} \pi_a  \pi_{\phi} - L  N {1 \over l^2} a  \phi
\cr
\sp(3.0)} \eqno(23)$$
Then the Hamilton equations of motion are
$$ \eqalign{ \sp(2.0)
a^{\prime} = {\partial H \over {\partial \pi_a}} = - {N \over L} 
\pi_{\phi},
\cr
\phi^{\prime} = {\partial H \over {\partial \pi_{\phi}}} = - {N \over L} 
\pi_a,
\cr
\pi_a ^{\prime} = - {\partial H \over {\partial a}} = L  N {1 
\over l^2} \phi,
\cr
\pi_{\phi} ^{\prime} = - {\partial H \over {\partial \phi}} = L  N {1 
\over l^2} a.
\cr
\sp(3.0)} \eqno(24)$$
In addition, varying (23) with respect to $N$ we learn that
$$ \eqalign{ \sp(2.0)
 - L \pi_a  \pi_{\phi} - L {1 \over l^2} a  \phi = 0.
\cr
\sp(3.0)} \eqno(25)$$
It is relatively straightforward to show that the solution that satisfies 
with (24) and (25) is the solution discovered by Ba$\tilde n$ados et al. 
[10] in the case of the spherically symmetric geometry under suitable 
boundary conditions. The canonical quantization is easily done by 
following the conventional procedure. After setting up the gauge 
conditions and the canonical commutation relations among $(\pi_a, a)$ 
and $(\pi_{\phi}, \phi)$, we need to solve an operator equation of the 
Hamiltonian constraint (25), that is, the Wheeler-DeWitt equation. 
Because of its simple structure, the Wheeler-DeWitt equation can be 
disentangled though we skip a detailed analysis of it at present.

Of course, the above discussion can be generalized to include various 
kinds of matter fields without difficulty. However, it might be not 
impossible but is very intricate to quantize a minisuperspace model 
such that the metric  
tensor depends on the general two dimensional coordinates $x^a$. This  
situation is in fact encountered in considering a dynamical 
black hole. This is our reason why we have constructed a rather 
general formalism. 

To make a quantization at all events, it seems to be natural to rely on 
some approximation method in order to solve the constraints 
analytically.  Though the usual approximation method must be a 
perturbation theory, one runs into problems of unrenormalizability in 
trying to quantize (2+1)-dimensional gravity in the metric formulation. 
Recently, however, Tomimatsu [6] proposed an interesting approximation 
method of constraints that has been applied for various problems by us 
[7-9]. His key idea is to solve the Hamiltonian and supermomentum 
constraints only in the vicinity of the apparent horizon of a black hole. 
In this paper, we will extend his idea to solve the constraints near the 
singularity.

To save a space, let us consider quantum gravity near the singularity 
$r =0$ and the apparent horizon $r = r_{+} \equiv l \sqrt M$ at the same time.  
Near 
the both places, it seems to be physically reasonable to assume that 
$$ \eqalign{ \sp(2.0)
\Phi \approx \Phi(v), \phi \approx r, M \approx M(v),
\cr
\sp(3.0)} \eqno(26)$$
which can be proved to be consistent with the equations of motion (4)-(6) 
[7]. Then, from (18) and (26) we learn
$$ \eqalign{ \sp(2.0)
\gamma &= M - {r^2 \over l^2},
\cr
&\approx \cases{ M(v) & for $r \approx 0$ \cr
                 0 & for $r \approx r_{+} = l \sqrt{M}$ \cr},
\cr
\sp(3.0)} \eqno(27)$$
and $\alpha = {1 + \gamma}$, $\beta = {1 \over \sqrt{\gamma}}$. And the 
canonical conjugate momenta are expressed by
$$ \eqalign{ \sp(2.0)
p_{\Phi} &\approx - {1 \over 2} \phi \partial_v \Phi,
\cr
p_{\phi} &\approx  - {1 \over {8 \gamma}} \partial_v M ,
\cr
p_{\gamma} &\approx -{1 \over 8}.
\cr
\sp(3.0)} \eqno(28)$$
It is remarkable that near the singularity and the apparent horizon the 
two constraints become proportional to each other
$$ \eqalign{ \sp(2.0)
- \gamma H_1 &\approx \sqrt{\gamma} H_0,
\cr
             &\approx {2 \over \phi} p_{\Phi}^2 + \gamma  p_\phi . 
\cr
\sp(3.0)} \eqno(29)$$
Here let us make some remarks. First of all, just at the apparent horizon 
$\gamma$ becomes zero, thus in order to avoid this singular behavior we 
assume  $\gamma$ to be  a small but finite constant [7]. This assumption 
is equivalent to the statement that we restrict ourselves to the 
interior near but not at the apparent horizon of a black hole. Of course, 
such an  
assumption is unnecessary in the vicinity of the singularity. Secondly,
the most interesting fact that is special to the (2+1)-dimensional
black hole is that 
we can construct quantum gravity even near the singularity. We can easily 
check that the present formulation makes no sense in trying to deal with 
curvature singularity existing as in the four dimensional black holes. 
Finally, it is also of interest to point out that the same forms of the 
constraint and the canonical conjugate momenta are shared near the 
singularity and the apparent horizon. The only difference exists in the 
value which $\gamma$ takes. 

Imposing the constraint (29) as an operator equation on the state, 
we have the well-known Wheeler-DeWitt equation 
$$ \eqalign{ \sp(2.0)
i {\partial \Psi \over {\partial \phi}} =  - {2 \over 
{\gamma \phi}} {\partial^2 \over 
{\partial \Phi^2}} \Psi,
\cr
\sp(3.0)} \eqno(30)$$
which we can rewrite as the Schr$\ddot o$dinger equation with the 
Hamiltonian $H = p_{\Phi}^2$ and the time $T = {2 \over \gamma} 
\log{\phi}$ in the superspace  
$$ \eqalign{ \sp(2.0)
i {\partial \Psi \over {\partial T}} = H  \Psi = p_\Phi ^2 \Psi.
\cr
\sp(3.0)} \eqno(31)$$

To find a special solution of this Wheeler-DeWitt 
equation, let us use the method of separation of variables. Then the 
solution is 
$$ \eqalign{ \sp(2.0)
\Psi = (B e^{- \sqrt A \Phi(v)} + C e^{ \sqrt A \Phi(v)} ) \ e^{ i {A 
 \over \gamma} T},
\cr
\sp(3.0)} \eqno(32)$$
where $A$, $B$, and $C$ are integration constants. Without losing a 
generality, let us take the boundary condition $C=0$. Under the 
definition of an expectation value $< \cal O >$ of an operator $\cal O$ 
$$ \eqalign{ \sp(2.0)
< {\cal O} > = {1 \over {\int d\Phi |\Psi|^2 }} \int d\Phi \Psi^* {\cal O} 
\Psi,
\cr
\sp(3.0)} \eqno(33)$$
it is straightforward to evaluate
$$ \eqalign{ \sp(2.0)
< \partial_v M > =  - {16 A \over <\phi>},
\cr
\sp(3.0)} \eqno(34)$$
where we have used the equation (28) or (29). Note that this result never 
depends on the value of $\gamma$, which is singular just at the apparent 
horizon. This result holds true near both the singularity and the 
apparent horizon.

Different choices of the integration constant $A$ yield different 
physical pictures. If one chooses the constant $A$ to be a negative 
constant, e.g., $- {1 \over 16} k_1 ^2$, this equation represents the 
absorption of the external matters by a black hole 
$$ \eqalign{ \sp(2.0)
< \partial_v M > = {k_1 ^2 \over <\phi>} ,
\cr
\sp(3.0)} \eqno(35)$$
for which the physical state has a form of the scalar wave propagating in 
black hole in the superspace
$$ \eqalign{ \sp(2.0)
\Psi = B e^{-i {1 \over 4} | k_1 | \Phi(v) - i {k_1 ^2 \over {16 
\gamma}} T}.
\cr
\sp(3.0)} \eqno(36)$$
On the other hand, if one takes the constant $A$ to be a positive 
constant, e.g., $ {1 \over 16} k_2 ^2$, (34) means the Hawking 
radiation [12] 
$$ \eqalign{ \sp(2.0)
< \partial_v M > = - {k_2 ^2 \over <\phi>},
\cr
\sp(3.0)} \eqno(37)$$
for which, as expected from the physical viewpoint, this time the 
physical state has an exponentially 
damping form in the classically forbidden region showing the quantum 
tunneling 
$$ \eqalign{ \sp(2.0)
\Psi = B e^{- {1 \over 4} | k_2 | \Phi(v) + i {k_2 ^2 \over {16 
\gamma}} T}.
\cr
\sp(3.0)} \eqno(38)$$

Now let us consider a physical situation where the neutral matters come 
in black hole across the horizon from the outside and approach the 
singularity. Our main concern is to investigate quantum behaviors of 
(2+1)-dimensional spherically symmetric black hole near the singularity 
in such a situation. The equation (35) tells us that as the matters approach 
the singularity the increase rate of black hole mass has a tendency to 
diverge infinitely. This phenomenon of an infinite divergence of the 
local mass function inside black hole under an incorporation of matters 
is somewhat similar to the mass inflation [13, 8] in the four dimensional 
Reissner-Nordstrom black hole although their relation is not clear at 
present. 

Nevertheless, let us proceed with this analogy further. In the mass 
inflation scenario,  
the exponential rise of the local mass function makes the inner Cauchy 
horizon unstable and 
change to the curvature singularity, as a result of which, it is 
prohibited that we  
travel to other universes with asymptotically flat spacetime regions via 
the charged wormhole.   What becomes of 
the case of (2+1)-dimensional black hole? This question may easily be 
answered by using the machinery that we have developed in this paper. Let 
us consider to evaluate an expectation value of the scalar curvature. 
From the equation of motion (4), it becomes  
$$ \eqalign{ \sp(2.0)
<  R  > = - {2 \over l^2}  + 2 <  g^{ab} \partial_a \Phi \partial_b 
\Phi  >. 
\cr
\sp(3.0)} \eqno(39)$$
On physical grounds, one expects quantum matter field 
$\Phi$  to become to have a dependency on not only $v$ but also $r$ 
variables through the strong quantum fluctuations near the singularity. 
Then the quantum field $\Phi$ can be expanded around the classical 
background $<  \Phi(v)  >$ as follows: 
$$ \eqalign{ \sp(2.0)
\Phi(v, r) \approx <  \Phi(v)  > + \delta \Phi(v, r),
\cr
\sp(3.0)} \eqno(40)$$
where $\delta \Phi(v, r)$ denotes the quantum fluctuation.
Now we can calculate the expectation value of the scalar curvature whose 
result is given by 
$$ \eqalign{ \sp(2.0)
<  R  > = - {2 \over l^2}  + { 2 | k_1 | \over {<\phi>}} \partial_r 
\delta \Phi,
\cr
\sp(3.0)} \eqno(41)$$
up to the leading approximation level with respect to $\delta \Phi$. 
Thus we arrive at an  
important conclusion that if there is a slight matter 
perturbation near the singularity from the outside the expectation value 
of the scalar curvature  
diverges at the singularity $r = 0$ in quantum gravity although the scalar 
curvature is finite classically there. It is worth mentioning here that 
$<  R  >$ is strictly finite at the apparent horizon as long as $\partial_r 
\delta \Phi|_{r=r_{+}} < + \infty$.

Finally, let us consider the case of the Hawking radiation (37) and (38). 
At first sight, one may be anxious that in this case $<  R  >$ also becomes 
the curvature singularity at the origin. However, an analogous 
calculation to (41) leads to
$$ \eqalign{ \sp(2.0)
<  R  > = - {2 \over l^2}  - i { 2 | k_2 | \over {<\phi>}} \partial_r 
\delta \Phi.
\cr
\sp(3.0)} \eqno(42)$$
Note that the second term in the right-hand side of this equation is 
purely an imaginary number in contrast with a real number in (41). 
Obviously, this result has relevance to a decaying character of 
evaporating  black hole, thus is harmless in the present formulation.

To summarize, in this article we have considered (2+1)-dimensional spherically 
symmetric black hole from a point of view of quantum instability of 
singularity. Once matters are incorporated into black hole, the curvature 
singularity is formed at the origin of black hole by the quantum effects. 
We believe that this conclusion is extremely general, and independent of 
any matter fields that we consider. It seems to be interesting to clarify the 
relation between the singularity in (2+1)-dimensional black hole and the 
inner Cauchy horizon in (3+1)-dimensional Reissner-Nordstrom black hole. 
In this respect, the works by Hiscock about evaporating 
black holes might be helpful [14]. We hope to return to this problem in 
near future.

\vskip 32pt
\leftline{\bf References}
\centerline{ } %
\par
\item{[1]} D.Page, in Proceedings of the Fifth Canadian Conference on 
General Relativity and Relativistic Astrophysics, edited by 
R.B.Mann et al. (World Scientific, Singapore, 1994). 

\item{[2]} K.S.Thorne, R.H.Price and D.A.Macdonald, Black Hole: The 
Membrane Paradigm (Yale University Press, 1986).

\item{[3]} G.'t Hooft, Nucl. Phys. {\bf B335} (1990) 138; Phys. Scripta 
{\bf T15} (1987) 143; ibid. {\bf T36} (1991) 247.

\item{[4]} I.Oda, Int. J. Mod. Phys. {\bf D1} (1992) 355; Phys. Lett. 
{\bf B338} (1994) 165; Mod. Phys. Lett. {\bf A10} (1995) 2775.

\item{[5]} B.S.DeWitt, Phys. Rev. {\bf 160} (1967) 1113. 

\item{[6]} A.Tomimatsu, Phys. Lett. {\bf B289} (1992) 283.

\item{[7]} A.Hosoya and I.Oda, ``Black Hole Radiation inside Apparent 
Horizon in Quantum Gravity'', Prog. Theor. Phys. (in press).

\item{[8]} I.Oda, ``Mass Inflation in Quantum Gravity'', EDO-EP-8, 
gr-qc/9701058.

\item{[9]} I.Oda, ``Evaporation of Three Dimensional Black Hole in 
Quantum Gravity'', EDO-EP-9.

\item{[10]} M.Ba$\tilde n$ados, C.Teitelboim, and J.Zanelli, Phys. Rev. 
Lett. {\bf 69} (1992) 1849; M.Ba$\tilde n$ados, M.Henneaux, C.Teitelboim, 
and J.Zanelli, Phys. Rev. {\bf D48} (1993) 1506.

\item{[11]} C.W.Misner, K.S.Thorne, and J.A.Wheeler, Gravitation (Freeman,
1973). 

\item{[12]} S.W.Hawking, Comm. Math. Phys. {\bf 43} (1975) 199.

\item{[13]} E.Poisson and W.Israel, Phys. Rev. {\bf D41} (1990) 1796.

\item{[14]} W.A.Hiscock, Phys. Rev. {\bf D23} (1981) 2813, 2823.

\endpage
%

%
\bye